\documentclass[12pt]{article}
\usepackage{times}
\usepackage{geometry}
\usepackage{graphicx}
\usepackage[dvipsnames]{xcolor}
\usepackage{natbib}
\usepackage{enumitem,amssymb}
\usepackage{ragged2e}
\newlist{thematic}{itemize}{8}
\setlist[thematic]{label=$\square$}
\usepackage{pifont}

\usepackage{hyperref}
\newcommand\myshade{100}
\colorlet{mylinkcolor}{violet}
\colorlet{mycitecolor}{YellowOrange}
\colorlet{myurlcolor}{Aquamarine}
\hypersetup{
  linkcolor  = mylinkcolor!\myshade!black,
  citecolor  = mycitecolor!\myshade!black,
  urlcolor   = myurlcolor!\myshade!black,
  colorlinks = true,
}
\usepackage[capitalize]{cleveref}

\begin{document}

\title{International Astrophysical Consortium \\for High-energy Calibration
\\
\bf{Summary of the 16th IACHEC Workshop}
}

\author{
C.~E.~Grant$^1$,
K.~K.~Madsen$^2$,
V.~Burwitz$^3$,
K.~Forster$^4$,
M.~Guainazzi$^5$, \and
V.L.~Kashyap$^6$, 
H.~L.~Marshall$^1$,
C.~B. Markwardt$^2$,
E.~D.~Miller$^1$, 
L.~Natalucci$^7$, \and
P.~P.~Plucinsky$^6$,
M.~Shidatsu$^{8}$,
Y.~Terada$^{9,10}$
}

\date{\today}   %do not edit this field!!

\newcommand{\artxc}{\textit{ART-XC}}
\newcommand{\erosita}{\textit{eROSITA}}
\newcommand{\rosat}{\textit{ROSAT}}
\newcommand{\xmm}{XMM-\textit{Newton}}
\newcommand{\chandra}{\textit{Chandra}}
\newcommand{\suzaku}{\textit{Suzaku}}
\newcommand{\swift}{\textit{Swift}}
\newcommand{\nicer}{\textit{NICER}}
\newcommand{\astrosat}{\textit{AstroSat}}
\newcommand{\nustar}{\textit{NuSTAR}}
\newcommand{\hxmt}{\textit{Insight-HXMT}}
\newcommand{\hitomi}{\textit{Hitomi}}
\newcommand{\xrism}{\textit{XRISM}}
\newcommand{\integral}{\textit{INTEGRAL}}
\newcommand{\fermi}{\textit{Fermi}}
\newcommand{\athena}{\textit{Athena}}
\newcommand{\ep}{\textit{Einstein Probe}}
\newcommand{\ixpe}{\textit{IXPE}}
\newcommand{\leia}{\textit{LEIA}}
\newcommand{\rxte}{\textit{RXTE}}
\newcommand{\EP}{\textit{Einstein Probe}}
\DeclareRobustCommand{\ion}[2]{\textup{#1\,\textsc{\lowercase{#2}}}}
\newcommand{\cstat}{{\tt c-stat}}
\newcommand{\etal}{\mbox{{\em et~al.}}}

\maketitle

{\centering
  $^1$Kavli Institute for Astrophysics and Space Research, \\ Massachusetts Institute of Technology, USA \\  
  $^2$NASA Goddard Space Flight Center, USA \\
  $^3$Max Planck Institute for Extraterrestrial Physics, Germany\\
  $^4$Cahill Center for Astronomy and Astrophysics, California Institute of Technology, USA \\

  $^5$ESA-ESTEC, The Netherlands \\
  $^6$Center for Astrophysics $|$ Harvard \& Smithsonian (CfA), USA \\
  $^7$IAPS-INAF, Italy \\
  $^8$Ehime University, Department of Physics, Japan \\
  $^9$Saitama University, Graduate School of Science and Engineering, Japan \\
  $^{10}$Japan Aerospace Exploration Agency, Institute of Space and Astronautical Science, Japan \\
}

\vspace{5mm}
\begin{center}
{\bf \large Abstract} \\
\end{center}

In this report we summarize the activities of the International Astronomical Consortium for High Energy Calibration (IACHEC) from the 16th IACHEC Workshop at Parador de La Granja, Spain. Sixty-one scientists directly involved in the calibration of operational and future high-energy missions gathered during 3.5 days to discuss the status of the cross-calibration between the current international complement of X-ray observatories, and the possibilities to improve it. This summary consists of reports from the Working Groups with topics ranging across: the identification and characterization of standard calibration sources, multi-observatory cross-calibration campaigns, appropriate and new statistical techniques, calibration of instruments and characterization of background, preservation of knowledge, and results for the benefit of the astronomical community.

\section{Meeting Summary}

The International Astronomical Consortium for High Energy Calibration (IACHEC)\footnote{\href{http://iachec.org}{\tt http://iachec.org}} is a group dedicated to supporting the cross-calibration of the scientific payload of high energy astrophysics missions with the ultimate goal of maximizing their scientific return. Its members are drawn from instrument teams, international and national space agencies, and other scientists with an interest in calibration. Representatives of over a dozen current and future missions regularly contribute to the IACHEC activities. 

IACHEC members cooperate within Working Groups (WGs) to define calibration standards and procedures. The objective of these groups is primarily a practical one: a set of data and results produced from a coordinated and standardized analysis of high-energy reference sources that are in the end published in refereed journals. Past, present, and future high-energy missions can use these results as a calibration reference. Table \ref{table:WGoverview} summarizes the WGs active during this report and their primary projects and areas of responsibility.

The 16th IACHEC meeting was hosted by Peter Kretschmar and Felix Fürst (both ESA-ESAC). It was held at Parador de La Granja in Spain, and the meeting was attended by sixty-one scientists representing high energy missions internationally. The meeting also featured presentations from IACHEC members unable to attend in person, who were able to view and contribute to the meeting via video teleconferencing. Advances in the understanding of the calibration of more than a dozen missions was discussed, covering multiple stages of operation and development:

\begin{itemize}
\item[-] Currently operating missions: \chandra, \ep, \erosita, \hxmt, \integral, \ixpe, \nicer, \nustar, \swift, \xmm, and \xrism\
\item[-] Pre-launch status: \textit{SVOM}
\item[-] Missions under development: \textit{eXTP}, \textit{NinjaSat}, and \textit{SMILE}
\item[-] Historical missions: Apollo 15/16 X-ray Fluorescence Spectrometer, \rxte 

\end{itemize}

In addition to various other topics, special emphasis was given at the general sessions of this workshop to presemtations on the calibration of the most recently launched missions: \xrism\ and \ep.

Alongside the contributed sessions, the IACHEC WGs held their meetings and this report summarizes the main results of the 16th meeting\footnote{The presentations are available at: \href{https://iachec.org/2024-parador-de-la-granja-spain}{\tt https://iachec.org/2024-parador-de-la-granja-spain}}, and comprises the reports from each of the IACHEC WGs: Calibration Uncertainties (\cref{s:calstat}), Detectors and Background (\cref{s:det_bkg}), Non-Thermal SNRs (\cref{s:nt_snr}), Thermal SNRs (\cref{s:t_snr}), Timing (\cref{s:timing}), and Isolated Neutron Stars and White Dwarfs (\cref{s:inswd}).

The IACHEC gratefully acknowledges sponsorship for the meeting from AHEAD 2020, the Society for Promotion of Space Science and the International Conference Support Program of Kanagawa Prefecture Japan, and from the \nustar, \swift, and \xmm\ missions.

The 17th IACHEC meeting will be held in Osaka Bay in Japan on 12-15 May 2025. More information about attending and presenting will be posted on the meeting web page.\footnote{\href{https://iachec.org/2025-osaka-bay-japan/}{\tt https://iachec.org/2025-osaka-bay-japan/}} 

\begin{table}[tbh]
    \centering
    \begin{tabular}{| p{0.27\linewidth}|p{0.21\linewidth}|p{0.45\linewidth}|}
        \hline
         Working Group & WG Chair & Projects  \\
         \hline
         \hline
         Calibration Statistics & Vinay Kashyap & Quantifying response uncertainties; statistical methods; Concordance project \\
         \hline
         Clusters of Galaxies & Eric Miller & Multi Mission Study of selected targets from the HiFLUGCS sample\\
         \hline
         Contamination & Herman Marshall & Definition, measurement, and mitigation of molecular contaminant \\
         \hline
         Coordinated Observations & Karl Forster & Organization of yearly cross-calibration campaign of 3c273; investigation of potential cross-calibration candidate 1ES 0229+200; publication of 3c273 cross-calibration campaign\\
         \hline
         Detectors and Background & Catherine Grant & Forum for discussion of detector effects; background measurement \& modeling\\
         \hline
         Heritage & Matteo Guianazzi & Curation of the IACHEC work; the IACHEC source database (\href{http://iachecdb.iaps.inaf.it}{\tt ISD})\\
         \hline
                 Non-thermal SNR & Lorenzo Natalucci \& Craig Markwardt & Cross-calibration with the Crab and G21.5-0.3\\
         \hline
         Thermal SNR & Paul Plucinsky & Cross-calibration with 1E 0102.2$-$7219 (E0102), N132D, and Cas~A; definition of standard models \\
         \hline
         Timing & Yukikatsu Terada \& Megumi Shidatsu & Summary of Timing performance and calibration status across missions; Systematic timing cross-calibration with Crab archive data; Planning timing cross-calibration campaign\\
         \hline
         White Dwarf and Isolated Neutron Stars & Vadim Burwitz & Cross-calibration with RX\,J1856.5$-$3754 and 1RXS\,J214303.7$+$065419 \\
         \hline
    \end{tabular}
    \caption{IACHEC Active Working Groups}
    \label{table:WGoverview}
\end{table}

\section{Working Group reports}

\subsection{Calibration Uncertainties: CalStats and High-Resolution Spectra}\label{s:calstat}

The Calibration Statistics (CalStats) Working Group focuses on methods and techniques that are applied to calibrating telescopes, and tools that are developed for that purpose.  Note that in usual astronomical parlance, the term {\sl calibration} is used to characterize the stationary measurement uncertainties (like PSFs and RMFs) and bias corrections (like ARFs and detector gains) imposed by the instrumentation, and incorporating them into analyses so that the measured signals have reliable physical interprations.  But in the field of statistics itself, the term has a different, more generalized meaning\footnote{\url{https://en.wikipedia.org/wiki/Calibration_(statistics)}}: it is a process of {\sl inverse regression}, where a predictive model is used to map an observation to the explanatory variable (that is, if $Y=mx+c$, then using a measurement of $Y$ to infer $x$ via the linear model [see, e.g., Osborne 1991]).

The main WG session was conducted jointly with the High-Resolution WG, and primarily involved a discussion by H.\ Marshall (MIT) of (1) polarimetry statistics (Marshall 2024), and (2) an update on the status of the Concordance Project (Chen et al.\ 2019, Marshall et al.\ 2021).  The former dealt with how variances in polarization estimates are underestimated if the polarization of the background is ignored, and how the error contours depend on the electric vector position angle (EVPA).  The latter described future enhancements like incorporating detailed passband information via empirically determined power-law spectral models, and requested inputs to update the $\tau$ matrix which defines the priors for the systematic uncertainties in each instrument.  The CalStats WG maintains a list of known $\tau$'s at its wiki site\footnote{\href{https://wikis.mit.edu/confluence/display/iachec/Calibration+Statistics\#CalibrationStatistics-Theτ-matrixofeffectiveareauncertainties}{The $\tau$-Matrix of Effective Area uncertainties}}.  Requests were made for input from several instrument teams that attended the IACHEC~XVI workshop (\ep\ FXT, WXT; \erosita; \ixpe; \nicer; {\sl NinjaSat}; {\sl SMILE} SXI; {\sl SVOM}; \xrism\ Resolve, Xtend) and the table will be updated as information is gathered.

The High-Resolution WG dealt with the measurement and cross-calibration of line velocities in the active star HR~1099, which was observed simultaneously with \xrism\ Resolve, reported by E.\ Miller (MIT) and \chandra\ ACIS-S/HETG, reported by V.\ Kashyap (SAO).  Several inconsistencies between the measurements were noted, and attributed mainly to different modeling methods by the two teams: the former relying on a globally defined correlation, and the latter relying on shifts in individual line measurements.

Several talks during IACHEC~XVI touched upon concepts relavant to CalStats during the workshop:\begin{itemize}
    \item Pileup modeling schemes for ACIS data using Bayesian hierarchical modeling and simulation based inference, and neural network empirical modeling (V.\ Kashyap (SAO))
    \item Parametric ARF and RMF modeling in \xmm\ and \erosita\ (K.\ Dennerl (MPE))
    \item Exploration of the causes of differences in cluster temperature estimates in hard bands of \chandra\ and \xmm\ compared to \nustar\ (D.\ Wik (University of Utah))
    \item Modeling of low counts spectra to understand effects of temperature dependent flux on exoplanets (S.\ Rukdee (MPE))
    \item A description of the \xrism\ Xtend Transient Search system (M.\ Yoshimoto (Osaka University))
    \item Discussion of the capabilities of \href{https://docs.stingray.science/en/stable/}{\tt Stingray} spectral+timing algorithms (M.\ Bachetti (INAF-Osservatorio Astronomico di Cagliari))
    \item On defining the background\footnote{Known information about background models in various instruments is kept in the \href{https://wikis.mit.edu/confluence/display/iachec/Calibration+Statistics\#CalibrationStatistics-Libraryofbackgroundmodels/scripts/packages}{CalStats Wiki Library of background models/scripts/packages}.} in \ixpe\ by (S.\ Silvestri (INFN-Pisa))
\end{itemize}

The CalStats WG has previously hosted several off-calendar talks and extended discussions\footnote{For example, a \href{https://hea-www.harvard.edu/AstroStat/CHASC_2324/index.html\#ycmb_20240228}{CHASC talk} on {\tt C-stat} goodness of fit and using it to characterize systematic errors, by Y.\ Chen and M.\ Bonamente in February 2024}, and plans to continue to schedule talks of interest as when opportunities arise, on topics from statistics methods to machine learning applications.

\subsection{Clusters of Galaxies}\label{s:clusters}

This WG uses clusters of galaxies as cross-calibration standard X-ray sources. Massive clusters have several advantages as calibration sources: the X-ray emission of the hot intracluster medium does not vary with time, it is bright across a broad energy band, and it has a fairly simple continuum-dominated spectrum. However, clusters are spatially extended and often contain bright, variable AGN, complicating comparison between instruments with very different imaging characteristic.  
Cluster science and the use of clusters as calibration sources are important for many existing and upcoming X-ray missions, including notably \erosita, \xrism, and \athena, and the WG has been active in the past (e.g., Nevalainen et al.\ 2010, Kettula et al.\ 2013, Schellenberger et al.\ 2015).

The Clusters WG meeting at the 16th IACHEC Workshop drew a small number of attendees who discussed making progress on the Multi-Mission Study (MMS), a project aiming to compare X-ray spectroscopic results of a sample of clusters obtained with current and past X-ray missions. This project has been ongoing for several years, and we have identified representatives among the WG membership for \xmm\ EPIC MOS and pn, \chandra\ ACIS, \suzaku\ XIS, \swift\ XRT, \nustar, \hxmt, \astrosat, \nicer, \rosat, and, new this year, \xrism. The representative for each instrument is tasked with gathering existing data spanning our cluster sample, applying the most recent calibration, extracting spectra and responses, and providing these to the WG chair for MMS cross analysis. This data will also be provided to the CalStats WG for inclusion in the Concordance effort (see Section \ref{s:calstat}). We expect to select a small number of clusters for an initial comparison, drawing from the HiFLUGCS sample (Reiprich \& B\"ohringer 2002) hot ($kT>6$ keV), nearby ($z<0.1$) systems with at least 100,000 counts accumulated in the central 6$^\prime$ in each instrument, observed no more than 3$^{\prime}$ off-axis. This study will form the bulk of the WG work during the foreseeable future.

\subsection{Detectors and Background}\label{s:det_bkg}

The Detectors and Background WG had one well attended session. The Detectors WG provides a forum for cross-mission discussion and comparison of detector-specific modeling and calibration issues, while the Background WG provides the same for measuring and modeling instrument backgrounds in the spatial, spectral and temporal dimensions.  Attendees represented many past, current, and future X-ray missions, including \athena, \chandra, \ep, \nustar, \xrism, and \xmm.  As existing missions go deeper, and planned missions get more ambitious, understanding and modeling background and detector response is all that much more important.

Our working group session had four talks, plus an extended discussion.  Claudio Pagani (University of Leicester) discussed the upcoming \textit{SMILE}-SXI CCD instrument and the tools being developed to model and mitigated the expected radiation damage effects. Nick Durham (SAO) presented the \chandra\ ACIS calibration effort to use the supernova remnant Cas A as a replacement for the decaying on-board radioactive source for gain tracking. Ivan Valtchanov (ESAC) presented an update to the long-term CTI calibration on \xmm\ EPIC-pn. Finally, Vinay Kashyap (SAO) reported on developing new techniques to handle pileup in \chandra\ ACIS observations.  Slides from the presentations are available on the IACHEC web page.

\subsection{Non-Thermal SNRs}\label{s:nt_snr}

The mission of the non-thermal SNR WG is to define the model of the two Pulsar Wind Nebula (PWN) as reference standards:  G21.5$-$0.9 (below 10 keV) and of the Crab.   

G21.5$-$0.9 is a plerionic pulsar wind nebula of a few mCrab, 3 arminutes across, and with no detectable pulsations, which makes it well suited as a calibration candle for a variety of instruments.  Since the IACHEC paper by Tsujimoto et al.\ (2011), which investigated the cross-calibration of G21.5-0.9 between \chandra, \integral, \rxte, \suzaku, \swift, and \xmm, there has not been a concerted effort to compare observations with newer observatories.  The group has proposed to start up a new investigation into defining a standard spectrum for G21.5$-$0.9.  However, in the past year, this effort has been of lower priority.

As the brightest non-thermal supernova remnant in the X-ray sky, the Crab Nebula remains the focus of calibration work by the group.   Although Wilson-Hodge et al.\ (2011) established that the Crab nebula's flux varies with time by about 10\%, considerable interest remains in using it for relative flux calibrations by comparing contemporaneous observations by multiple observatories.

To this end, L.\ Natalucci (INAF-IAPS) continues work on a multi-epoch Crab paper.  This effort identifies time epochs of contemporaneous or nearly-contemporaneous X-ray observations for the purposes of flux comparison in four energy bands covering the 3--300 keV range.  To date, the team has collected data from the \xmm\ EPIC, \suzaku\ XIS, GSO, \& PIN,  \swift\ BAT, \rxte\ PCA, \integral\ ISGRI \& SPI, \nustar, \fermi\ GBM, and \astrosat\ CZTI instruments.  With additional observatories coming online in recent times, the group proposes to add epochs in 2017 and 2020 to allow contributions from \hxmt\ ME \& HE, and \nicer. 

Continuing the topic of hard X-ray measurements of the Crab, J.\ Rodi (INAF-IAPS) presented work on observations of the Crab by \integral\ PICsIT, in the 300--2000~keV energy range. This effort represents the first real foray into understanding the PICsIT flux calibration after \integral's launch.  In its current state, PICsIT analysis retrieves Crab fluxes in that energy band that are approximately 50\% lower than those reported by SPI during the same time interval (Jourdain \& Roques 2020), but comparable to those reported by COMPTEL (Kuiper et al.\ 2001).  It is expected that PICsIT fluxes would be revised upward by $\sim$30\%, but this does not explain the total difference between it and SPI.  Work continues to more fully understand PICsIT calibration. 

The Crab spectrum has spectral curvature, gradually advancing the power law photon index from about 2.10 to 2.15 over the energy range of 80--120~keV.   This has led to differences between published spectral slopes that are energy dependent.  Individual instrumental flux measurements over a limited energy range can also be impacted because of this issue.   Recent efforts led by E.\ Jourdain (IRAP) have shown that a GRBM (log-powerlaw) type of model is a better fit across the 1--150~keV bandpass.  In addition, C.\ Markwardt (NASA/GSFC) presented efforts tied to \nicer\ X-ray spectroscopy that attempts to assimilate several previous works into a usable reference model for Crab spectral fitting over a broad range of X-rays and low energy gamma rays.  These include the work of Kaastra et al. (2009) dedicated to \xmm\ observations for X-ray absorption, dust and nebular shape; and the work of Kuiper et al. (2001) for broad band SEDs.  Drawing upon these efforts, the group aims to establish a form of reference spectral model for both the Crab nebula and pulsar.

\subsection{Thermal SNRs}\label{s:t_snr}\label{s:n132d}

The thermal supernova remnants (SNRs) WG aims to use the line-rich spectra of 1E 0102.2-7219 (E0102), N132D, and Cas A to improve the response models of the various
instruments (gain, CTI correction, QE, spectral redistribution function, etc) and
to compare the absolute effective areas of the instruments at the energies of the
bright line complexes.  The spectra of E0102 and N132D are assumed to be time
invariant on the scale of decades, as opposed to Cas A that shows significant
spectral variations on a decade timescale.  Our efforts focus on developing standard
models that can be used by the various teams to meaningfully compare their results.
The intention is that current and future missions will be able to use these standard
models to improve their calibrations.  The group met remotely once in May 2024
before the IACHEC meeting to coordinate our efforts and prepare for the meeting.

The thermal SNRs WG met once during the IACHEC meeting itself.  Paul Plucinsky (SAO)  
started off with a presentation of preliminary results from the \xrism\, Resolve
observations of N132D and Cas A.  The Resolve results on SNRs are highly anticipated
because they are the highest resolution spectra yet acquired for these objects and
are certain to produce new insights to inform the development of standard spectral
models. Unfortunately the Gate Valve on the Aperture
Assembly for Resolve has not opened at this point in the mission, so the effective
bandpass for Resolve is approximately 1.7 to 12.0 keV.  Both N132D and Cas A have
sufficiently hard spectra that useful results have been obtained for energies above
1.7 keV.  The Resolve spectra of N132D cleanly separate the \ion{Si}{XIII}~He$\alpha$~triplet
from the \ion{Si}{XIV}~Ly$\alpha$ lines and the \ion{S}{XV}~He$\alpha$~triplet
from the \ion{S}{XVI}~Ly$\alpha$ lines.  The Resolve spectra even separate the
forbidden and recombination lines of the \ion{Si}{XIII}~He$\alpha$ and
\ion{S}{XV}~He$\alpha$ triplets.  The Resolve spectra show that the Si and
S lines are relatively narrow and the \ion{Fe}{XXV}~He$\alpha$ triplet is quite
broad such that the forbidden, intercombination, and resonance lines are not resolved
even though the resolution of the instrument is sufficient to do so.  
The redshift of the Fe lines is significantly higher than that of the Si \& S lines (see
Audard et al.\ 2024), confirming an earlier result with \hitomi.
\xrism\, observed two locations in Cas A, one in the southeastern (SE) part and
one in the northwestern (NW) part of the remnant.  The spectra exhibit strong
lines of Si, S, Ar, Ca, Fe, and Ni and weak lines from Ti \& Cr and the odd-Z elements of P, Cl, K \& Mn. These spectra show the
redshift/blueshift structure from NW to SE known from CCD spectra but to higher
precision.  The spectra show that the H-like and He-like lines of Si and S have
different redshifts, a result that was not possible to obtain with CCD resolution spectra.
The broadening of the Si, S and Fe lines varies significantly with position in Cas
A, with the highest broadening toward the center and the lowest broadening at the
edges.  All of these new results should inform a standard spectral model for Cas A
in the future.

Nick Durham (SAO) presented results on developing a standard Cas A model that  he is
using to improve the ACIS calibration on \chandra.  He has generated an empirical
model with 48 Gaussians for the lines, a ``No-Line" APEC model for the thermal continuum,
and a powerlaw for the nonthermal continuum. He has used this model to track gain
and normalization changes at the energies of the bright lines as a
function of time.  He has also developed a simpler model with just five Gaussians
for the bright Si and S lines which shows the expected redshift/blueshift structure
of Cas A clearly.  Craig Markwardt (NASA/GSFC) showed his efforts to develop a
standard model that started with Andy Beardmore's (University of Leicester) model with three
bremsstrahlung components, Gaussians for the lines, and a powerlaw component. He
substituted an {\tt srcut} model for the nonthermal continuum in place of the powerlaw
component. He applied this model
to \nicer, \xmm\, EPIC pn, \xmm\, EPIC MOS, \swift\, XRT, \rxte\, PCA, \nustar, and \swift\, BAT and
compared the residuals.  There are $\pm20\%$ variations in the overall constant
applied to the model spectra for the different instruments.  He notes that a
challenge for this work is accounting for the different fields-of-views and
off-axis responses of the instruments given the significant spatial variations in the
Cas A spectra.

Tomokage Yoneyama (Chuo University) showed the \xrism\, Xtend spectra of E0102 collected
so far in the mission.  Xtend has a relatively large effective area $<2.0$~keV as it is
not affected by the Gate Valve issue.  Multiple Xtend observations of E0102 show no
change in the observed fluxes from the O \& Ne  lines and can therefore put an upper limit on
any possible contamination layer.  Konrad Dennerl (MPE) briefed the group on his
efforts to use the well-calibrated spectrum of E0102 to improve the \erosita\, and
\xmm\, EPIC-pn response products.  It has been challenging to develop a response matrix
for the pn that improves the fits for both line-rich and continuum sources. He has
developed a matrix which significantly improved the fits to the E0102 data from the
pn.  The line normalizations for the bright lines in the E0102 spectrum are
generally consistent with the IACHEC standard model values but there are variations
on the order of $\pm5\%$ over the 45 pn observations that span the 24 year \xmm\,
mission, however the global normalization appears to be $3\%$ lower than the IACHEC value.
The \erosita\, CCDs have a narrower spectral redistribution function than the pn CCDs
at low
energy such that the shapes of the O \& Ne lines more closely resemble a Gaussian with a less
pronounced shoulder at low energy. The global normalization for
\erosita\, also shows a $3\%$ difference compared to the IACHEC value.  Finally
Dennerl showed results on monitoring the flux of SN1987A from Chandreyee
Maitra (MPE) with \erosita\, and \xmm\, pn. The flux in the 0.5-2.0~keV band is
clearly declining but the flux in the 3.0--10.0 keV band appears to be increasing over the
last five years.

Adam Foster (SAO) updated the group on his efforts to create a standard model for
N132D in the high energy bandpass 4.5--8.0 keV based on \xmm\, MOS and pn data. The
model for the \ion{Fe}{XXV}~He$\alpha$ triplet is in good shape and should be
published soon. The model also provides a useful constraint on the strength of the
 \ion{Ca}{XIX}~He$\alpha$ triplet.  Martin Stuhlinger (ESAC) updated the group on his
 efforts to develop an empirical model for N132D in the low energy bandpass from
 0.3--1.5~keV based on \xmm\, RGS data.  There is a physical model published by
 Suzuki et al. 2020 but for calibration purposes it would be useful to have an
 empirical model. Martin has developed such a model with 108 Gaussians and four
 ``No-Line" APEC models for the continuum components. There are some interesting
 discrepancies in the continuum level between the model and the data at energies in between
 some of the bright lines.  It is not clear if this is truly continuum or
 unidentified lines. The long term goal is to merge these efforts at low energy
 and high energy to create a unified model for N132D.
 
\subsection{Timing}\label{s:timing}

 The Timing WG aims to provide a forum for in-orbit and on-ground timing-calibrations of X-ray missions, focusing on their timing systems, calibration methods, issues, and lessons learned. 
 The WG also aims to coordinate simultaneous observations of pulsars for timing calibrations with multi-X-ray missions and/or radio observatories.

We had three oral presentations related to the timing WG in the plenary session: ``Ground and In-orbit Verifications of the \xrism\ Timing System" by Y.\ Terada (Saitama University/JAXA-ISAS) and ``Stingray: Spectral Timing for All" by M.\ Bachetti (INAF-Osservatorio Astronomico di Cagliarai) and ``Timing calibration of the CubeSat X-ray observatory \textit{NinjaSat}" by N.\ Ota (RIKEN/Tokyo University of Science).
In the working-group parallel session, we had three other oral presentations; ``Pulsar Cross-calibration" by M.\ Bachetti, ``Challenges to Keep the Timing Accuracy of \xrism\ Timing System in GPS Failure Mode" by M.\ Shidatsu (Ehime University), and ``Timing Cross-calibration with the Crab in March 2024" by Y.\ Terada.
In summary, reports on timing calibrations of the newly born missions in 2023 (\xrism\ and \textit{NinjaSat}), and the tools and status of the systematic timing cross-calibration study of multiple missions were presented and discussed. 

The following three items are currently active in the Timing Working Group.
\begin{enumerate}
    \item Systematic timing cross calibration study
    \item Analysis of simultaneous timing calibration campaign of Crab in March 2024
    \item Summary table of the timing calibrations of X-ray missions
\end{enumerate}

\subsubsection{Systematic Timing Cross Calibration Study}

The goal of the first item is to perform the systematic study of timing accuracy among multiple missions. 
We continue this activity, in which X-ray missions such as \xmm, \suzaku, \nustar, \astrosat, \hitomi, and \swift\ participated.
New missions, \xrism, \textit{NinjaSat}, \ep, and \hxmt\ will soon join this activity this year.
The details of the updates of the Python tool, Stingray, were presented.
The keypoint of this study is to use the single code and the same ephemeris for the consistency in TOA calculation among multiple datasets. 
The required inputs from instrument members of the mission: dead-time-corrected event FITS files with HEASARC-type timing keywords, and the information of the barycentric-correction sky position.
Then, we will get the list of arrival times from multiple missions in various epochs, and the difference between them. 
As a result, we see several outliers in several missions. 
The details of the method and the current result will be summarized in the near-future paper.

\subsubsection{Timing Calibration Campaign of Crab in March 2024}

In the second item, we performed the two sets of simultaneous Crab observations in March 2024 as an IACHEC activity; one in early March with \ep, \xmm, \integral, and \nustar, and the other in the middle of March with \xrism, \nicer, \nustar, and \textit{NinjaSat}.
The preliminary results of the Crab pulse profile from the second campaign were presented. 
The short-term action items for the analyzes were summarized; 1) finishing the timing calibration for \xrism, 2) applying the latest clock correction table and dead-time correction for \nustar, and 3) search for proper screening criteria on this observation for \nicer.
The results will be published as either the \xrism\ or IACHEC timing calibration papers soon.

In relation to the first and second activities, we briefly discussed the barycentric correction tool in the HEAsoft package; {\it barycen} for \suzaku, \hitomi, and \xrism\ and {\it barycorr} for \nicer, \nustar, etc.
Crab timing calibration only requires the solar system ephemeris JPL-DE200, because the Jodrell-Bank radio ephemeris uses this system; however, solar system ephemeris JPL-DE420 is recommended in other cases. 
We, the IACHEC timing WG, decided to send a request to the HEASARC support JPL-DE420 on {\it barycen}.

\subsubsection{Summary table of the timing calibrations of X-ray missions}

As reported in the last IACHEC meeting in 2023, we have summarized calibration requirements and the status in one table. The table is available on the IACHEC Wiki page.\footnote{\href{https://wikis.mit.edu/confluence/display/iachec/Timing}{\tt https://wikis.mit.edu/confluence/display/iachec/Timing}}
In this IACHEC meeting, we have no major update on this activity, but we decided to include new missions, \xrism\ and \ep.

\subsection{Isolated Neutron Stars and White Dwarfs}\label{s:inswd}

This WG aims to improve the cross-calibration of X-ray telescopes in the low energy range ($<$ 1.5 keV) by using spectra of Isolated Neutron Stars (INSs) and White Dwarfs (WDs). These objects should not display time-dependent variation and have physically well-modeled spectra that can be used as spectral standard candles at low energies. Over the years, a set of white dwarfs (GD153, Sirius B, and HZ43) with spectra that can be described by physical white dwarf models, and the isolated neutron star RX\,J1856.5$-$3754 (RX\,J1856) with a spectrum that can be best be described by a single black-body model, have met the characteristics of a standard candle. The three WDs and RX\,J1856  were used to improve the calibration of the low energy end of the \chandra\ LETGS (LETG+HRC-S) and provide a cross-calibration with \rosat\ and recently also with \nicer\  and \erosita, and the \EP\ Follow-up X-ray Telescope. In the context of \erosita\ a new INS is being studied and used for calibration: 1RXS\,J214303.7$+$065419 (RX\,J2143). 

V.~Burwitz (MPE) gave an overview of the status of the INSs and WDs working group and pointed out that standard RX\,J1856 blackbody model presented on the wiki page has been updated to include the most recent \chandra/LETGS observations. 

V.\ Kashyap (SAO) presented the results of using HZ43 for the \chandra\ HRC calibration on behalf of P. Ratzlaff, B. Wargelin, and J. Drake. HZ43 is a DAwk white dwarf at 60 pc; it is an extremely soft X-ray source with most emission at $\lambda$\,$>$\,60\,{\AA} (E\,$<$\,0.2\,keV) $\lesssim$\,2.5 ct/s in HRC-I/LETG0, $\lesssim$\,6.5\,ct/s in HRC-S/LETG0. It is used for the QE (HRC-I) and QEU (HRC-S) updates at low E, for gain calibration for HRC-I and HRC-S, and for PSF monitoring on both HRC-I and HRC-S. It is observed at a cadence of at least once a year, plus every time there is a planned change in detector characteristics (like voltage changes). The measurements of the flux vs. time with respect to the beginning of the mission were shown. This trend in this data was used to create a correction function for the HRC-I/LETG0, HRC-S/LETG0, and HRC-S/LETG+1/-1.

H.\ Marshall (MIT) gave an update on the  \chandra/ACIS contaminant. This calibration is done using \chandra\ LETG + ACIS observations of RX\,J1856. The contaminant deposition rate continues increasing above the 2015-20 linear trend; the existing model looks good and does not need to be updated. RX\,J1856 is nearing the limit of being useful to monitor this contaminant deposition trend. The source is still detectable in 60 ks (5.4~$\sigma$), not including extended soft X-ray ﬂux. The contaminant correction is low by 35\% in 2023.

K.\ Dennerl (MPE) presented the \xmm/EPIC-pn and \erosita\ RMF and ARF improvements that can be obtained using the standard candle model for RX\,J1856 available on the IACHEC wiki page. For \erosita, it is possible to track contamination after larger orbital maneuvers, but fortunately, no cumulative effect is detectable. The \erosita\
RXJ\,1856 spectrum can be fit using the IACHEC model with only the normalization free. After adjusting the ARF and RMF, good spectral fits can be obtained for ``TM8" \& all valid patterns.

C.\ Markwardt (NASA/GSFC) presented an update of \nicer\ observations of the isolated NS RX\,J1856 and the white dwarf HZ43. He used the baseline IACHEC model for RX\,J1856 and the Beuermann et al. 2006 model for HZ43. An easier-to-use model of the HZ 43 spectrum, based on the publicly available TMAP white dwarf atmosphere model, has been constructed, added to the IACHEC wiki page, and presented. Results of show that the NICER detected flux at 0.28\,keV (HZ 43) and 0.35\,keV (RXJ\,1856) is low by $\sim$16\% and $\sim$14\% respectively. This could be caused by additional absorption, either an error of $~$20\% in the absorption column or differences in the material properties of the filters; these were measured before launch.

In summary, the INS RX\,J1856.5$-$3754 and white dwarf HZ43 are being observed and used by many X-ray observatories for calibration purposes to monitor the status of their low energy calibration.

\section*{References\footnote{see \href{https://iachec.org/papers/}{\tt https://iachec.org/papers/} for a complete list of IACHEC papers}}

\noindent
Audard, M., et al.\ 2024, \href{https://doi.org/10.1093/pasj/psae080}{PASJ, 76, 1186}\\
\noindent
Beuermann, K., Burwitz, V., Rauch, T. 2006, 
\href{https://doi.org/10.1051/0004-6361:20065478}{A\&A 458, 541}\\
\noindent
Chen, Y., et al.\ 2019, \href{https://doi.org/10.1080/01621459.2018.1528978}{J.Am.Stat.Assoc., 114:527, 1018}\\
\noindent
Jourdain, E.\ \& Roques, J. P.\ 2020, \href{https://doi.org/10.3847/1538-4357/aba8a4}{ApJ, 899, 131}\\
\noindent
Kaastra, J., et al.\ 2009, \href{https://doi.org/10.1051/0004-6361/20077892}{A\&A, 497, 291}\\
\noindent
Kettula, K., Nevalainen, J., \& Miller, E.D.\ 2013, \href{https://doi.org/10.1051/0004-6361/201220408}{A\&A, 552, A47}\\
\noindent
Kuiper, L., et al.\ 2001, \href{https://doi.org/10.1051/0004-6361:20011256}{A\&A, 378, 918}\\
\noindent
Marshall, H., et al.\ 2021, \href{https://doi.org/10.3847/1538-3881/ac230a}{AJ, 162, 254} \\
\noindent
Marshall, H.\ 2024, \href{https://doi.org/10.3847/1538-4357/ad0897}{ApJ, 964, 88} \\
\noindent
Nevalainen, J., David, L., \& Guainazzi, M.\ 2010, \href{https://doi.org/10.1051/0004-6361/201015176}{A\&A, 523, A22}\\
\noindent
Osborne, C.\ 1991, \href{https://doi.org/10.2307/1403690}{International Statistical Review, 59, 309}\\
\noindent
Reiprich, T.H. \& B\"ohringer, H.\ 2002, \href{https://doi.org/10.1086/338753}{ApJ, 567, 716}\\
\noindent
Schellenberger, G., et al.\ 2015, \href{https://doi.org/10.1051/0004-6361/201424085}{A\&A, 575, A30}\\
\noindent
Suzuki, H., et al.\ 2020, \href{https://doi.org/10.3847/1538-4357/aba524}{ApJ, 900, 39}\\
\noindent
Tsujimoto, M., et al.\ 2011, \href{https://doi.org/10.1051/0004-6361/201015597}{A\&A, 525, 25} \\
\noindent
Wilson-Hodge, C., et al.\ 2011, \href{https://doi.org/10.1088/2041-8205/727/2/L40}{ApJ, 727, 40}\\
\\

\end{document}